\documentclass[%
 reprint,
 amsmath,amssymb,
 aps,nofootinbib,   
]{revtex4-1}

\usepackage{bm}% bold math
\PassOptionsToPackage{linktocpage}{hyperref}
\usepackage[hyperindex,breaklinks]{hyperref}
\usepackage{enumitem}
\usepackage{slashed}

\usepackage{xcolor}
\usepackage{float}

\usepackage{array}
\usepackage{mathtools}

\usepackage{etoolbox}

\begin{document} 

\title{Gravity, Superselection Rules and Axions}
 %from de Sitter Quantum Breaking?}

\author{  C\'esar G\'omez} 
\affiliation{Instituto de F\'{i}sica Te\'orica UAM-CSIC, Universidad Aut\'onoma de Madrid, Cantoblanco, 28049 Madrid, Spain}
\affiliation{
	Arnold Sommerfeld Center, Ludwig-Maximilians-Universit\"at, Theresienstra{\ss}e 37, 80333 M\"unchen, Germany
}
\affiliation{
	Max-Planck-Institut f\"ur Physik, F\"ohringer Ring 6, 80805 M\"unchen, Germany
}
%\affiliation{ 
%	$^c$Center for Cosmology and Particle Physics, Department of Physics, New York University, 726 Broadway, New York, NY 10003, USA
%}

%\date{\today}

\begin{abstract}
It is generally accepted that consistent coupling to quantum gravity implies that any low energy free parameter is in reality the vacuum expectation value of some local quantum field. In this note we present a modest attempt to prove this general claim using as  key example the $\theta$ parameter of QCD and the Peccei-Quinn axion solution of the strong CP problem. The gravitational emergence of the local axion field associated with the superselection $\theta$ parameter is worked out using as basic ingredients the quantum representation of classical geometries in terms of coherent states and the interpretation of the $\theta$ QCD superselection rule in quantum information terms as the lack of a quantum reference frame for topological charge. The formal connection with the model independent axion in string theory is briefly discussed.

\end{abstract}
\maketitle

%\pacs{(old) 14.60.Pq,13.15.+g,04.60.-m,11.30.Rd}

\section{Introduction}

One of the main lessons of string theory, as a theory of quantum gravity, is that in quantum gravity there are not free parameters and that all of them should be interpreted as expectation values of some dynamical fields. 

The most striking consequence of this basic principle is that even those parameters representing would be superselection charges are vev's of some associated field. The most canonical example of this state of facts is the Peccei Quinn (PQ) solution \cite{PQ}, \cite{Weinberg},\cite{Wilczek} of the strong CP problem where the $\theta$ parameter, originally introduced as a superselection charge, is promoted, after introducing the axion, into its vev. This leads us to the following very general question: How consistent coupling to quantum gravity transforms superselection charges into vev's of new fields?

This is the question we shall try to address in this note. In order to identify the way quantum gravity can be used to generate the local axion field and consequently to evade the rol of  $\theta$ QCD as a superselection (SS) parameter we shall use two basic ingredients. On one side the interpretation, pioneered in \cite{susskind}, of superselection rules as the lack of a reference frame to define some observables ( see \cite{RF} for a review ) and on the other side the quantum representation of space time geometry, put forward in \cite{us1,us2,us3,Zell}, as a {\it coherent state} of gravitons \footnote{This coherent state approach to geometry contrast with the much more popular vision where the quantum version of geometry is modeled using a {\it thermal state} ( see for instance \cite{GH1,GH2, Malda,Raamsdonk}). We will make some brief comments on the relevance of this difference for the problem under study.}.

Our main finding can be summarized as follows. Associated with the $\theta$ parameter of QCD there exists a canonically {\it conjugated} observable (the topological number ) that requires in order to be defined of a reference frame. The lack of such reference frame is what makes the $\theta$ parameter a SS charge. The PQ global charge defines this conjugated observable. Consistent coupling to gravity promotes this couple of conjugated global observables into the conjugated variables $\phi, \pi$ of the local axion field. In string theory, as we shall briefly discuss at the end of this note, the two {\it conjugated} variables are naturally introduced through the Bianchi identity underlying the anomaly cancellation \cite{GS} mechanism. In summary it seems that gravity abhors global SS charges and that naturally completes the theory with the addition of its conjugated local observable \footnote{The Kalb Ramond field in the string case.}. Hence it looks that a basic principle of quantum gravity is that {\it if $A$ belongs to the algebra of physical observables $\cal{A}$ there exists in $\cal{A}$ its canonically conjugated observable} \footnote{Very likely this principle implies the quantum gravity inconsistency of global symmetries.}. 

\section{Superselection rules}
In the mathematical foundations of quantum mechanics as stablished by Von Neumann \cite{VN} it was assumed, not only that any physical observable is represented by a self adjoint operator, but also the opposite, namely that any self adjoint operator represents a physical observable. It was Wigner ( see \cite{W} ) the first to consider the possibility that the space of physical observables is smaller than the total space of self adjoint operators. This possibility leads to the formal definition of superselection rules (SSR). Indeed a superselection charge is an operator $\hat Q$ that conmutes with the whole algebra ${\cal{A}}$ of physical observables
\begin{equation}
[\hat Q, {\cal{A}}] =0
\end{equation}

 The existence of such operator has important consequences. The obvious one is that the physical Hilbert space decomposes into a direct sum of Hilbert spaces each one characterized by a given eigenvalue of $\hat Q$. But the most surprising one is that quantum superpositions of states having different eigenvalues of $\hat Q$ cannot be physically realized. In more physical terms what this implies is that such quantum coherent superposition requires, in order to be prepared, infinite amount of energy. 
 
% For the simplest example of the $\theta$ parameter of Yang Mills we can imagine the different $\theta$ sectors as being separated by an infinite potential barrier. 

A more basic case of what would be a superselection rule and the one originally used in \cite{WWW}  is the electric charge. The heuristic way \cite{SW} to understand why in this case $\hat Q_e$ should conmute with the whole algebra ${\cal{A}}$ of physical observables is to reduce those to local observables with support in the interior of space and to represent the operator $\hat Q_e$, using the operator representation of the Gauss law, as the integral of the electric field, over the asymptotic region of space. In this approximation standard locality will imply that the so defined algebra ${\cal{A}}$ of physical observables conmutes with $\hat Q_e$. In this formal argument we leave out of the algebra of physical observables those located in the asymptotic region and consequently the SSR associated with the electric charge depends on the former assumption of the un-observability of those asymptotic data.  The first attempt to define coherent local superpositions of states with different electric charge was worked out in \cite{susskind}. In this approach coherent states $|Q_e,\phi\rangle$  with $\phi$ the {\it canonically conjugated variable} to the electric charge were used to prepare locally coherent superpositions of states with different electric charge.

This first discussion makes clear that the notion of SSR strongly depends on what we previously accept as an observable. In other words, once we accept that some self adjoint operator is unobservable we create a SSR and consequently also the corresponding superselection parameter. This fact provides the hint that will drive us in our discussion. Indeed if quantum gravity eliminates a would be SS parameter in favor of the vev of some new field what in essence we are learning is that consistent coupling to quantum gravity demands to enlarge the algebra of observables. 

Before further elaborating these ideas let us extract some obvious consequences of the existence of a SSR and its connection with {\it classicality}. Let us think in a couple of {\it conjugated} variables satisfying $[A,B]=\hbar$. Obviously if $A$ is a SS charge then by definition $B$ cannot be in the algebra of physical observables. This fact has a simple but deep consequence, namely if the conjugated variable $B$ is out of the algebra of observables then we are forced to conclude that whenever we measure $A$ we do it with zero quantum uncertainty and that un-observability of $B$ is due to the complementary fact, namely that whenever we try to measure $B$ we get infinite uncertainty. This makes the observable playing the role of SS actually classical since in all observations, independent on what conditions, its value is known with infinite precision. Is this intrinsic classicality what kills the quantum coherence of superpositions of states with different eigenvalues of the SS charge $A$. 

In other words we cannot built quantum superpositions of eigenvectors with different values of the SS charge because {\it secretly its conjugated variable is unobservable, meaning that we cannot prepare any experimental device that could deliver its value with finite uncertainty.}

\section{Observability, reference frame and Superselection}

In many cases the observability and consequently the definition of an observable and the corresponding quantum state is only possible relative to some {\it reference frame} \footnote{For the reference frame approach to SSR see \cite{RF} and references therein. For the potential use of SSR in quantum protocols see for instance \cite{cirac} \cite{kitaev}}. Moreover, changes of reference frame define unitary transformations acting on the Hilbert space. If we define physical states as those invariant under change of reference frame the physical Hilbert space is associated with an irrep of the group of reference frame transformations. The parameters characterizing these irreps play de role of SS charges. 

In this sense we can say that a given observable defines a SS charge if 

{\it a) the conjugated variable requires in order to be defined the choice of a reference frame and 

b) we lack any operative definition of reference frame that we can share among different observers. }

The crucial element is the point b) above. Many observables require to set a reference frame. This is needed already to define position. What in general we do is to associate reference frame with the definition of an observer and to use the change of reference frame as the way to relate the descriptions of the {\it same physics} by two different observers. This is the basis of relativity principles. However, the point b) above refers to those cases where the reference frame used by one observer, let us say Alice, is totally unknown by the other observer that we shall denote Bob. In these conditions Bob is forced to describe, the state that Alice is preparing, as the sum over all the possible states associated with the whole set of reference frames that Alice could be using. In other words, for Bob, the particular observable that Alice is measuring, using her reference frame, is not observable and consequently his description of Alice pure state, in case this state is a quantum superposition between different eigenvectors of the corresponding observable, is given by a density matrix with full decoherence. 

The only pure states available to Bob, describing Alice preparations, are those invariant under changes of Alice reference frame and will be characterized by a SS charge. In addition Bob will never see quantum superpositions of states with different values of the SS charge. 

In more precise terms let us denote $\hat Q$ the generator of the transformation defining a change of reference frame. We define this transformation as  $U_{\alpha} = e^{i\hat Q \alpha}$ with $\alpha$ representing what we can generically think as the {\it relative angle} between the different reference frames. We shall reduce our discussion to reference frame transformations defining a $U(1)$ group. Now let us denote $|\psi\rangle$ the state prepared by Alice using her reference frame. If Bob has no knowledge on how Alice reference frame is related to his reference frame his description of the state $|\psi\rangle$ will be given by 
\begin{equation}\label{mixed}
\int d(\alpha) U_{\alpha}|\psi\rangle\langle\psi|U_{\alpha}^{\dagger}
\end{equation}
where we integrate over the whole group with the appropriated Haar measure. Once we represent the group of reference frame transformations, the Hilbert space decomposes into a direct sum of irreps each one labeled by the eigenvalue of $\hat Q$. Thus the only states that Bob can use to describe Alice's states are those invariant under the action of $U_{\alpha}$ i.e. pure states with well defined value of $\hat Q$ or mixed states, like (\ref{mixed}), with full decoherence relative to the observable $\hat Q$. In other words, whenever Bob lacks information about the reference frame needed to define a given observable he needs to use the corresponding generator $\hat Q$ as a SS charge and to identify his physical Hilbert space with those states invariant under changes of reference frame.

Let us now make the former discussion a bit richer. The states invariant under $U_{\alpha}$ are those satisfying $\hat Q|\psi\rangle = Q|\psi\rangle$ for some value of the SS parameter $Q$. Let us now introduce formally another $U(1)$ generator that we shall denote $\hat Q_{t}$ satisfying
\begin{equation}
[\hat Q, \hat Q_{t}] = 1
\end{equation}
where from now on we shall take $\hbar=1$.
We can think of $\hat Q_{t}$ as the {\it conjugated} to the observable $\hat Q$.
Let us now define a basis of eigenvectors of $\hat Q_{t}$ 
\begin{equation}
\hat Q_{t}|\beta\rangle = \beta |\beta\rangle
\end{equation} 
From the commutation relations between $\hat Q$ and $\hat Q_{t}$ we get
\begin{equation}\label{change}
e^{i\hat Q \alpha}|\beta\rangle = |\beta +\alpha\rangle
\end{equation}
Thus we can define the invariant state $|Q\rangle$, satisfying $e^{i\hat Q \alpha}|Q\rangle = e^{i\alpha Q}|Q\rangle$ as
\begin{equation}
|Q\rangle = \int d(\beta) e^{iQ\beta} |\beta\rangle
\end{equation}
Moreover we easily find that
\begin{equation}
e^{i\hat Q_{t} \alpha}|Q\rangle = |Q+\alpha\rangle
\end{equation}
The former relations make explicit the meaning of the conjugated operator $\hat Q_{t}$. Indeed if $\hat Q$ is a SS charge then $\hat Q_{t}$ cannot be part of the algebra of physical observables. Moreover the conjugated eigenvalue $\beta$ is the one that changes under change of reference frame as can be seen from (\ref{change}). In other words, is the observable for which we don't have a reference frame. The way we deal with the lack of reference frame for this observable is to promote the conjugated one into a SSR. We will now use this formal frame to deconstruct the origin and meaning of the axion field.

\section{How to promote a SS charge into a vev? the axion recipe}
\subsection{The $\theta$ vacua}
Let us briefly recall the introduction of the axion field in QCD. In pure Yang Mills with gauge group $G$ we can naturally define using the temporal gauge $A_0=0$ the topological number $n$ as $\int d^3x K$ for $K$ the 3-form defined by $dK=F\wedge F$. This allows us to define perturbative vacua as associated with pure gauge configurations with a given value of this topological charge i.e. $|n\rangle$ \cite{CDG},\cite{JR}. The 3-form $K$ is not gauge invariant under topologically non trivial gauge transformations in $\Pi_3(G)$. In particular if $g$ is a gauge transformation with winding number one in $\Pi_3(G)$ the corresponding action on the Hilbert space of states is given by $U_g|n\rangle = |n+1\rangle$. Hence, and in order to define vacua which are invariant under these transformations, we need to use coherent superpositions of the $|n\rangle $ vacua where we integrate, with the appropriated Haar measure, over all the topologically non trivial gauge transformations. This defines the well known \cite{CDG1},\cite{JR} $\theta$ vacua as
\begin{equation}
|\theta\rangle = \sum_n e^{i\theta n} |n\rangle
\end{equation}
with the $\theta$ parameter playing the role of a SS charge. 

The path integral representation of the vacuum to vacuum amplitude $\langle \theta|\theta\rangle$ induces the extra term in the lagrangian $\theta F\wedge F$ that breaks CP. The PQ solution \cite{PQ} to the strong CP problem requires to add a new axion field $\phi(x)$ with i) invariance under PQ shift transformations $\phi(x)\rightarrow \phi(x) + c$ and such that ii) the PQ Noether current $J_{PQ}$ generating these transformations should be anomalous $\partial J_{PQ} = F\wedge F$. These two ingredients set the physical value of the $\theta$ angle as the vev of the axion \footnote{The main target in this note is to understand the logic of the axion generation and not its phenomenology so we shall make explicit in the discussion the contribution to the effective $\bar \theta$ from the quantum flavor dynamics.}.

\subsection{Topology and reference frame}
Let us now repeat the same story but from the point of view of observability and reference frame. In order to do that we define the operator $\hat Q_{\theta}$ satisfying $\hat Q_{\theta}|\theta\rangle =\theta|\theta\rangle$. Thus $\hat Q_{\theta}$ is the generator of the non trivial gauge transformations
\begin{equation}
e^{i\hat Q_{\theta} \alpha}|n\rangle = |n+\alpha\rangle
\end{equation}
with here $\alpha$ the integer parametrizing the topological number of the transformation. In other words $\hat Q_{\theta}$ is the generator of changes of the reference frame used to define the topological number $n$.

Following the discussion in the former section we formally introduce another conjugated charge $\hat Q_{t}$ that measures the topological number $n$ i.e.
\begin{equation}
\hat Q_{t} |n\rangle = n|n\rangle
\end{equation}
Note now that
\begin{equation}\label{one}
e^{i\hat Q_{t} \beta} |\theta\rangle = |\theta+\beta\rangle
\end{equation}
and that
\begin{equation}
[\hat Q_{\theta}, \hat Q_t] =1
\end{equation}
Now we can reach some interesting conclusions:
\begin{itemize}
\item If $\hat Q_{\theta}$ is a SS charge i.e. if the QCD $\theta$ parameter is a SS parameter then the conjugated $\hat Q_{t}$ is unobservable and out of the algebra of physical observables. In other words, $\theta$ is a SS parameter because {\it we lack any reference frame to measure the topological number} $n$. Moreover the introduction of the $\theta$ angle as a SS parameter in Yang Mills simply reflects the lack of a reference frame for measuring the topological number $n$ in $\Pi_3(G)$.
\item Adding the generator $\hat Q_{t}$ is equivalent to add the extra $U(1)$ PQ symmetry as can be observed from (\ref{one}). We shall denote this generator the PQ generator.
\item Including the PQ generator in the algebra of observables is equivalent to provide a reference frame for the topological number $n$. 

\item The PQ current formally associated with $\hat Q_{t}$ is anomalous {\it by construction}. The reason is simply because 
\begin{equation}
\Delta Q_{t} = n (+) - n(-) = \int F\wedge F
\end{equation}
Thus this PQ current satisfies all the properties we expect for the standard PQ symmetry. 
\end{itemize}

What about the axion ? Once we use the couple of conjugated generators $\hat Q$ and $\hat Q_{t}$ we can associate with these couple of conjugated operators the corresponding algebra of creation annihilation operators and to  build the {\it zero mode part} of the axion field $\phi(x)$ in terms of them. To do that let us think of an scalar field $\phi(x)$ with shift symmetry and let us denote the classical degenerate vacua $|a\rangle$. The generator of the shift symmetry is given in terms of the zero mode creation annihilation operators by $a_0-a_0^{+}$. Now identify this generator with $\hat Q_{t}$ defined above. In this way you get {\it the zero mode part $\phi$ of the axion field} from the creation annihilation operators associated with the couple of conjugated operators $\hat Q_{\theta}, \hat Q_{t}$ as well as $\hat Q_{t}$ as the generator of the PQ shift symmetry $\partial_0\phi$ and $\hat Q_{\theta}$ as $\phi$.

{\it Remarks}

1) Until now we have ignored in the discussion the rol of fermions. In case we have massless fermions chiral transformations can be used to transform the $\theta$ parameter. In this case the generator of the chiral transformation of the fermions defines the charge $\hat Q_{t}$ and consequently eliminates the role of $\hat Q$ as SS charge. The PQ current is now replaced by the chiral current $J_5$.

2) In the previous formal discussion we have omitted two crucial parameters. One is the mass scale $f_a$ needed to define the axion field $a(x)$ with appropriated dimensions and secondly the coefficient $\chi$ in the anomaly $\Delta Q_{t}= \chi \int F\wedge F$. This second coefficient was set in our construction without fermions equal to one. This changes if we couple the axion field to fermions.

3) The key ingredient of the PQ mechanism is the anomaly of the PQ current. This was automatic in our construction where the PQ generator is identified with $\hat Q_{t}$. Indeed in this case $\Delta Q_t$ is equal to the difference $n(+)-n(-)$ of the topological number $n$ in $\pm\infty$ and this difference is given by $\int F\wedge F$ by the definition of $n$.

\subsection{The Etiology of Axions}
In the general frame we are considering the origin of axions can be understood as follows:
\begin{itemize}
\item Assume you have a SS charge $\hat Q_A$ defining the generator of some reference frame  set of transformations and assume that this group of transformations is $U(1)$. 
\item Define the corresponding conjugated charge $\hat Q_B$ with $[\hat Q_A,\hat Q_B] = 1$. That the charge $\hat Q_A$ is a SS charge means that $\hat Q_B$ is unobservable i.e. that we lack a reference frame to measure this observable. In other words if $\hat Q_A$ is a SS then $\hat Q_B$ is the observable for which we lack a reference frame and $\hat Q_A$ is the generator of transformations between different reference frames. 
\item $\hat Q_B$ is the generator of the corresponding PQ symmetry. Since both $\hat Q_A$ and $\hat Q_B$ are global conjugated charges we can only associate with them {\it one} algebra of creation annihilation operators. The zero mode part of the corresponding axion field, let us say $\phi$ is defined using as creation annihilation operators the ones naturally defined by the couple of conjugated operators $\hat Q_A, \hat Q_B$ with $Q_A=\phi$ and $Q_B=\partial_0\phi$. 

\item If we enlarge the algebra of creation annihilation operators to the whole set of $a_k,a^{+}_k$ we can define the local axion field $\phi(x)$. Adding this field is what allows us to prepare {\it local} coherent superpositions of states with different value of the eigenvalue of $\hat Q_A$. 
\end{itemize}
The properties of this axion field depend on the dynamics. If we denote $|m\rangle$ the eigenvectors of $\hat Q_{B}$ the amplitude of the process
\begin{equation}
\langle m|m'\rangle 
\end{equation}
will set the anomalous conservation of $\hat Q_{B}$.  For non abelian gauge theories invariant under the group $G$, the charge $m$ is simply the topological number in $\Pi_3(G)$ and $\Delta \hat Q_B$ is given by the Pontryaguin number. 

What we lack at this point, once we have introduced the conjugated global charge, is to identify the mechanism to generate the $a_k,a^{+}_k$ algebras of operators. This will be our next task and the place where coupling to gravity will become crucial. 

\section{Energy and SS charges: $E(\theta)$}

Coming back to the case of $\theta$ in QCD and once we have defined the conjugated charge $\hat Q_t$ to the SS charge $\hat Q_{\theta}$  we can introduce the creation annihilation operators $a$ and $a^{+}$ canonically associated with this couple of conjugated global charges. Using {\it only} this algebra of zero mode operators we can define coherent states $| Q_{\theta}, Q_t\rangle \equiv |\theta, \phi\rangle$ where the coherent state label is defined as $\alpha = \theta e^{i\phi}$. This state is given by
\begin{equation}\label{coherent}
|\theta, \phi\rangle = e^{-N_0/2} \sum \frac{(N_0(\theta)e^{2i\phi})^{n/2}}{\sqrt{n!}} |n\rangle
\end{equation}
Defining as usual time by $\phi/m$ where we introduce a reference frequency $m$ we get that the Hamiltonian defining the time evolution of this state is given by $\hat N$, the number operator $\hat N|n\rangle =n|n\rangle$, and that the density of energy is given by
\begin{equation}\label{energy}
E(\theta) = \frac{m <\hat N>}{V} = \theta^2 f_a^2 m^2
\end{equation}
for $N_0(\theta)\equiv Vm f_a^2\theta^2$.
Note that this energy density is simply derived as the energy density of a {\it coherent state defined by the algebra of creation annihilation operators associated with the couple of operators $\hat Q_{\theta}$ and $\hat Q_{t}$}. After defining the PQ generator as $\hat Q_{t}$ these creation annihilation operators create axionic zero modes ($k=0$ ) with mass $m$. In addition to the frequency $m$ we have added the scale $f_a$ in order to define a density of energy. The meaning of this scale will be discussed in a moment.

Now we can compare (\ref{energy}) with what we normally assume is the case for pure Yang Mills \cite{Witten1} \cite{Witten2},\cite{Witten3} namely, for small $\theta$, 
\begin{equation}\label{top}
E(\theta) = C \theta^2
\end{equation}
with $C$ given by the topological susceptibility. This leads to the well known definition of the axion mass
\begin{equation}\label{energy2}
m^2 = \frac{C}{f_a^2}
\end{equation}
that not surprisingly is the analog of Witten Veneziano formula \cite{Veneziano} for the $\eta'$ mass if we replace $f_a$ by $f_{\pi}$.

In other words the logic path we have presented can be summarized as follow:

{\it Once we introduce the conjugated PQ charge $\hat Q_t$ and the corresponding algebra of zero modes creation annihilation operators $a$ and $a^{+}$ the $\theta$ vacua is promoted into a coherent state with the expected dependence of the energy density on $\theta$.}

{\it Remarks}

1) Note that in all this construction we have only used the zero mode part i.e. only {\it one} set of  creation annihilation operators.

2) A completely different representation of the $\theta$ vacua is as a  dilute gas  of instantons \cite{CDG2}. This representation leads to a density of energy $E(\theta)$  exponentially suppressed like $E(\theta)_{inst} = D cos(\theta) e^{-\frac{8\pi^2}{g^2(\rho)}}$ with $D$ representing the instanton scale measure factor $\frac{d\rho}{\rho^5} (\frac{8\pi^2}{g^2(\rho)})^{2N}$. This energy density is actually not well defined without adding some large scale cutoff on instanton size integration. In what follows we shall not discuss the instanton contribution to $E(\theta)$. This extra contribution to the vacuum energy can be relevant to find the minima for $\theta$ within the experimental window.

In summary once we have added to the algebra of observables the conjugated PQ global charge $\hat Q_{t}$, the $\theta$ vacua are promoted into coherent states with well defined energy density. In other words, the $\theta$ vacuum energy density already is telling us indirectly that we have added the conjugated charge $\hat Q_{t}$.

\section{Gravity and Superselection}
Once we have discovered $E(\theta)$ gravity enters into action. Until now what we have obtained including the PQ generator $\hat Q_{t}$ is to have the necessary structure to define the {\it zero mode} part of the algebra of creation annihilation operators $a$ and $a^{+}$ that we can think as the $a_0$ and $a_0^{+}$ part of the standard algebra of creation annihilation operators of a scalar field. Thus, in order to claim that we have a true axion field, {\it we need to generate the modes $a_k$ and $a_{k}^{+}$}. Next we shall offer a heuristic way to see how these modes naturally emerge once quantum gravity effects are taken into account.

First let us note that the existence of a finite density of energy $E(\theta)$ allows us to continue thinking in $\theta$ as a SS parameter. Indeed $\theta$ stops being a SS parameter if we can define {\it finite quantum amplitudes between states with different value of $\theta$}. This is equivalent to say that we can define local quantum superpositions between states with different values of $\theta$. Even if different $\theta$ vacua, defined as coherent states created with the zero mode part algebra $a,a^{+}$, come with different density of energy, the corresponding quantum state has infinite total energy in the limit of infinite volume. Thus the amplitude between states with different value of $\theta$ goes to zero in the infinite volume limit. In these conditions is perfectly fine to think in different SS sectors with different vacuum energy densities since the amplitude to transit from one to another is zero. 

To transit from one value of $\theta$ to a different one should be done {\it locally} and requires to transform the difference in the local density of energy into the creation of quanta with some non vanishing momentum $k$ i.e. {\it requires to have the whole algebra of creation annihilation operators $a_k,a_k^{+}$}. In other words, to have local transitions between different values of $\theta$ is only possible if in addition to $a_{0},a_{0}^{+}$ we have the whole set of $a_k,a_k^{+}$ or equivalently we have generated the local axion field. So the question we face now is: How to generate these extra modes ? or equivalently: How to generate the full fledged axion field and not only its zero mode part ?

In what follows we shall argue that quantum effects of gravity do this job. Indeed in presence of gravity each value of $E(\theta)$ leads to a different de Sitter space time with different cosmological constant
\begin{equation}
\Lambda(\theta) = \frac{C\theta^2}{M_P^2}
\end{equation}
Now let us model, following \cite{us1}, the corresponding Hubble patch as a coherent state of gravitons with typical momentum $k= \frac{\sqrt{C} \theta}{M_P}$. Due to the universality of gravity, these gravitons can now transfer a momentum to the axion zero mode $a_{0},a^{+}_{0}$
and to transform them into $a_k,a_{k}^{+}$ modes. This will happen due to non vanishing transfer momentum amplitudes described formally as
\begin{equation}\label{Amplitude}
axion (k=0) + graviton(k) \rightarrow axion (k_1) + graviton(k_2)
\end{equation}
with $k_1+k_2 =k$ and where the graviton transfers a momentum $k_1$ to the axion zero mode. If these sort of amplitudes are {\it different from zero} them the whole set of creation annihilation operators $a_k,a^{+}_k$ will be generated. Reciprocally if this amplitude is zero for any non vanishing momentum transfer then we will find that the different $\theta$ states continue working as SS sectors and that coherent local superpositions between different values of $\theta$ cannot be constructed. 

In the coherent state approach approach \cite{us1} the amplitude (\ref{Amplitude}) can be estimated and scales like $\frac{1}{N}$ for $N= \frac{R_H^2}{L_P^2}$ for $R_H$ the corresponding Hubble radius. In case of the de Sitter generated by $E(\theta)$ we get
\begin{equation}
N= \frac{M_P^4}{C\theta^2}
\end{equation}
We observe that the amplitude only vanishes if $N=\infty$. This happens for $\theta=0$ meaning that this value represents a stable vacuum in agreement with what we expect from \cite{VW}. Thus, any other value of $\theta$ in presence of gravity delivers the $a_k,a^{+}_k$ modes we need to induce local transitions between the different values of $\theta$. 

In summary coupling to gravity provides, through processes like (\ref{Amplitude}) a natural mechanism to generate the full fledged local axion field and to evade the SS properties of $\theta$. However this only takes place if (\ref{Amplitude}) is non vanishing or equivalently if we treat de Sitter fully quantum mechanically and not semi classically.

It could be worth at this point to make very explicit the steps of the deconstruction mechanism of the axion and to highlight the point where gravity enters. The mechanism has three steps.
\begin{itemize}
\item In the first step we add to the would be SS charge $\hat Q_{\theta}$ the conjugated PQ global generator $\hat Q_t$. Using these two operators we define the {\it zero mode} algebra of creation annihilation operators $a,a^{+}$.
\item In the second step, and using this zero mode part of the algebra, we define the $\theta$ vacua as coherent states and compute the energy density $E(\theta)$. In this second step we need to introduce a reference frequency $m$.
\item Finally in the third step, and using the gravitational back reaction to this energy as a de Sitter space we generate, using the amplitude (\ref{Amplitude}) the algebra of $a_k,a_k^{+}$ for $k$ non vanishing. These operators allows us to define the local axion field $a(x)$ and to define transitions between different values of $\theta$. Only in this third step we evade locally the SS nature of $\theta$. Moreover only in this last step we identify the equation of motion of $a(x)$ and the relation of the mass $m$ and the topological susceptibility.
\end{itemize}

{\it Remarks}

1) The amplitude (\ref{Amplitude}) was computed using the CSG representation of the quantum de Sitter state. In this sense the third step above crucially depends on the {\it quantum gravity} representation of de Sitter. Indeed in the semiclassical limit where $N=\infty$ i.e. where de Sitter is treated classically we will be unable to use the same argument to generate the local axion field i.e. the operators $a_k,a_k^{+}$. Hence semiclassical de Sitter is not enough to dynamically generate the local axion field.

2) Here we see a key difference between the coherent state approach and the most popular approach based on a {\it thermal state}. If we think in the de Sitter quantum state as a thermal state what we will get is instead of the coherent state (\ref{coherent}) the mixed state
\begin{equation}
\sum e^{\beta n}\frac{(N_0e^{2i\phi})^{n}}{\sqrt{n!}} |n\rangle\langle n|
\end{equation}
for $\beta$ the corresponding dS temperature. Thus, at least superficially the TSG approach is not enough to generate the local axion field.

3) Once the axion field is generated the anomaly for $\hat Q_t$ leads to the coupling $a(x) F\wedge F$ and to the equation of motion $\partial \partial a = \langle a|F\wedge F |a\rangle $.

4) The de Sitter coherent state we use to create locally quantum superpositions of different values of $\theta$ recalls the definition of reference frame used by Aharonov and Susskind in \cite{susskind}. The corresponding reference frame, applied to our case, is designed as a finite cavity where a coherent state for the couple of operators $\hat Q_{\theta},\hat Q_t$ is defined. The coherent state describing the Hubble patch of de Sitter associated with $E(\theta)$ plays naturally this role.

\subsection{The gravitational meaning of $f_a$ and the string model independent axion.}
The axion decay constant $f_a$ has been introduced in order to create an axion field with canonical normalization. Once we define the axion field the meaning of this constant is to set  the amplitude to create quanta with momentum $k$. This amplitude scales as $\frac{k^2}{f_a^2}$. In our construction we have assumed that the amplitude to create modes with $k$ non vanishing comes from gravitational transfer momentum amplitudes as (\ref{Amplitude}). This means that in this approach where the generation of the axion field is gravitational
\begin{equation}
\frac{k}{f_a} \sim \frac{k}{M_P}
\end{equation}
making $f_a$ naturally of the order of the Planck scale. This leads to a small axion mass $m^2 \sim \frac{\Lambda_{gauge}^4}{M_P^2}$ where we define the topological susceptibility of the gauge part defining the $\theta$ parameter
as $\Lambda_{gauge}^4$. 

It is now instructive to compare what we have done in order to get the axion field with the emergence of the model independent axion in string theory \cite{WittenO},\cite{WittenW} where the value of $f_a$ is also naturally given by the Planck scale \cite{Banks}\cite{Wittenaxion}. In string theory the axion field $\phi$ can be defined using the 3-form $H$ and the Bianchi identity $dH= tr R\wedge R- tr F\wedge F$. The fact that $H$ is globally defined allows us to define the PQ axion in terms of the Hodge dual $\star H$
\begin{equation}
Y_{\mu} = \frac{1}{f_a} \partial_{\mu}\phi
\end{equation}
with $Y^{\mu} = \frac{1}{6} \epsilon^{\mu\nu\alpha\beta}H_{\nu\alpha\beta}$. Now the term in the lagrangian $H\wedge \star H$ generates the kinetic term for the model independent axion field $\phi$. 

The formal connection with our construction is as follows. The topological number $n$ is defined by the 3-form $K$, so adding the PQ charge $Q_t$ is essentially equivalent to add to $K$ an {\it exact} 3-form\footnote{In string theory this is done through the exact component $dB$ of $H$ in 
\begin{equation}
H=dB -3\omega_{Y} + 3\omega_{L}
\end{equation}
 for $\omega$'s the gauge and gravity 
Chern-Simons forms.}. Moreover the generation of the full fledged axion field that we have motivated using gravity is the analog of the existence of the 3-form $H$ in string theory i.e. of a globally defined 3-form with the corresponding term in the lagrangian $H\wedge \star H$ and naturally interacting and transferring momentum with gravitons. 

Moreover in our frame the definition of the PQ charge $\hat Q_t$ as the {\it conjugated} to $\hat Q_{\theta}$ makes formally this charge the Hodge dual of a globally defined 3-form. Its existence reflects the conjugation between $Q_{\theta}$ and $Q_t$ and it reveals a deep connection between evading $\theta$ as a SS parameter and the very stringy generation of the Kalb Ramond 3-form $H$.

\section{Some comments on the Cosmological Constant} 
A natural question, in the present framework, is if the cosmological constant (cc) understood as a real constant and {\it in pure gravity} is playing the natural role of a SS charge. If we parametrize the cc $\Lambda$ by the Gibbons Hawking number $N= \frac{M_P^2}{\Lambda}$ the question is if in {\it pure gravity} this number plays the role of a SS parameter. Following our approach the answer will require to look for a candidate to conjugated variable and to see if this conjugated variable is unobservable or better to say if it is a variable for which we lack a natural reference frame. The simplest possibility for the conjugated variable is obviously some variable representing  $\dot N$ with
\begin{equation}
[\hat N,\dot {\hat N}] =1
\end{equation}
 and to think that the cc is a SS parameter if we {\it lack a clock} to define $\dot N$. 

One of the outputs that the CSG approach delivers is a natural {\it quantum clock } defining $\dot N$ \cite{us1}. Now assuming that we can try to proceed as we did above, we can use this couple of operators to generate the algebra of creation annihilation operators $a$ and $a^{+}$ as well as the whole tower with non vanishing $k$ using a similar mechanism based on gravitational transfer momentum. We know now what are these operators, simply the creation annihilation operators defining the inflaton field in the spirit presented in \cite{us1}. So it looks that in the CSG approach  the cc  naturally comes with its conjugated clock and  cannot be a SS charge i.e. a constant, parameter. In essence the eternity of de Sitter is the manifestation of the cc as a SS parameter and reciprocally if de Sitter is {\it not eternal} the cc is not a SS parameter and the former construction follows leading to the existence of new fields driving the mechanism of local quantum superpositions i.e. the mechanism of inflation. This line of reasoning  can be summarized saying that the cc {\it is a SSR if we lack a reference cosmological clock}. Hence, once consistent quantum gravity provides the need of such a clock the cc stops being a constant and consequently an eternal SSR. 

{\bf Acknowledgements.}
 I would like to thank Gia Dvali and Sebastian Zell for comments and many enlightening and relevant discussions. This work was supported  by the ERC Advanced Grant 339169 "Selfcompletion" and  the grants SEV-2016-0597, FPA2015-65480-P and PGC2018-095976-B-C21. 


\begin{thebibliography}{10}

   \bibitem{PQ}
  R.~D.~Peccei and H.~R.~Quinn,
  ``CP Conservation in the Presence of Instantons,''
  Phys.\ Rev.\ Lett.\  {\bf 38} (1977) 1440.
  doi:10.1103/PhysRevLett.38.1440
  \bibitem{Weinberg}
  S.~Weinberg,
  ``A New Light Boson?,''
  Phys.\ Rev.\ Lett.\  {\bf 40} (1978) 223.
  doi:10.1103/PhysRevLett.40.223
  \bibitem{Wilczek}
  F.~Wilczek,
  ``Problem of Strong  $P$  and  $T$  Invariance in the Presence of Instantons,''
  Phys.\ Rev.\ Lett.\  {\bf 40} (1978) 279.
  doi:10.1103/PhysRevLett.40.279
   \bibitem{susskind}
  Y. Aharonov and L. Susskind
  "Charge superselection rule"
Phys. Rev. 155, 1428 (1967)

  \bibitem{RF}
S. D. Bartlett, T. Rudolph, and R. W. Spekkens
"Reference frames, superselection rules, and quantum information"
Rev. Mod. Phys. 79, 555 (2007)
  \bibitem{us1}
  G.~Dvali and C.~Gomez,
  ``Quantum Compositeness of Gravity: Black Holes, AdS and Inflation,''
  JCAP {\bf 1401} (2014) 023
  doi:10.1088/1475-7516/2014/01/023
  [arXiv:1312.4795 [hep-th]].
  
  \bibitem{us2}
  G.~Dvali and C.~Gomez,
  ``Quantum Exclusion of Positive Cosmological Constant?,''
  Annalen Phys.\  {\bf 528} (2016) 68
  doi:10.1002/andp.201500216
  [arXiv:1412.8077 [hep-th]].
  
  \bibitem{us3}
  G.~Dvali, C.~Gomez and S.~Zell,
  ``Quantum Break-Time of de Sitter,''
  JCAP {\bf 1706} (2017) 028
  doi:10.1088/1475-7516/2017/06/028
  [arXiv:1701.08776 [hep-th]].
   \bibitem{Zell}
  G.~Dvali, C.~Gomez and S.~Zell,
  ``A Proof of the Axion?,''
  arXiv:1811.03079 [hep-th].
  \bibitem{GH1}
  G.~W.~Gibbons and S.~W.~Hawking,
  ``Action Integrals and Partition Functions in Quantum Gravity,''
  Phys.\ Rev.\ D {\bf 15} (1977) 2752.
  doi:10.1103/PhysRevD.15.2752
  
  \bibitem{GH2}
  G.~W.~Gibbons and S.~W.~Hawking,
  ``Cosmological Event Horizons, Thermodynamics, and Particle Creation,''
  Phys.\ Rev.\ D {\bf 15} (1977) 2738.
  doi:10.1103/PhysRevD.15.2738
  
  \bibitem{Malda}
  J.~M.~Maldacena,
  ``Eternal black holes in anti-de Sitter,''
  JHEP {\bf 0304} (2003) 021
  doi:10.1088/1126-6708/2003/04/021
  [hep-th/0106112].
  \bibitem{Raamsdonk}
  M.~Van Raamsdonk,
  ``Building up spacetime with quantum entanglement,''
  Gen.\ Rel.\ Grav.\  {\bf 42} (2010) 2323
   [Int.\ J.\ Mod.\ Phys.\ D {\bf 19} (2010) 2429]
  doi:10.1007/s10714-010-1034-0, 10.1142/S0218271810018529
  [arXiv:1005.3035 [hep-th]].
  
  \bibitem{GS}
  M.~B.~Green and J.~H.~Schwarz,
  ``Anomaly Cancellation in Supersymmetric D=10 Gauge Theory and Superstring Theory,''
  Phys.\ Lett.\  {\bf 149B} (1984) 117.
  doi:10.1016/0370-2693(84)91565-X
    \bibitem{VN}
  J.V.Neumann "Mathematical Foundations of Quantum Mechanics" Princeton University Press 1955
   \bibitem{W}
 A. S. Wightman
Il Nuovo Cimento B Volume 110, Issue 5?6, pp 751?769
 \bibitem{WWW}

G. -C. Wick, A. S. Wightman, and E. P. Wigner
"Superselection Rule for Charge"
Phys. Rev. D 1, 3267 (1970) 

 \bibitem{SW}
 A. S. Wightman
"Proof of the charge superselection rule in local relativistic quantum field theory"
Journal of Mathematical Physics 15, 2198 (1974);
% \bibitem{Dirac}
% P.A.M.Dirac Proc.Roy.Soc A114,243 (1927)
%  \bibitem{Nieto}
%  P. Carruthers and M. M. Nieto
%  "Phase and Angle Variables in Quantum Mechanics"
%
%Rev. Mod. Phys. 40, 411 (1968)
%  \bibitem{Nieto2}
%  P. Carruthers and M. M. Nieto
%  "Coherent States and the Number-Phase Uncertainty Relation"
%P. Carruthers and M. M. Nieto
%Phys. Rev. Lett. 14, 387 (1965)
%

%\bibitem{RF}
%S. D. Bartlett, T. Rudolph, and R. W. Spekkens
%"Reference frames, superselection rules, and quantum information"
%Rev. Mod. Phys. 79, 555 (2007)
\bibitem{cirac}
 F. Verstraete and J. I. Cirac, "Quantum-nonlocality in the presence of superselection rules and some applications" quant-ph/0302039 (2003).
 
 \bibitem{kitaev}
 A.Kitaev, D. Mayers, J. Preskill
 "Superselection rules and quantum protocols"

Phys.Rev. A69 (2004) 052326
   \bibitem{CDG1}
  C.~G.~Callan, Jr., R.~F.~Dashen and D.~J.~Gross,
  ``The Structure of the Gauge Theory Vacuum,''
  Phys.\ Lett.\ B {\bf 63} (1976) 334
   [Phys.\ Lett.\  {\bf 63B} (1976) 334].
  doi:10.1016/0370-2693(76)90277-X
  \bibitem{JR}
  R.~Jackiw and C.~Rebbi,
  ``Vacuum Periodicity in a Yang-Mills Quantum Theory,''
  Phys.\ Rev.\ Lett.\  {\bf 37} (1976) 172.
  doi:10.1103/PhysRevLett.37.172
  \bibitem{Witten1}
  E.~Witten,
  ``Instantons, the Quark Model, and the 1/n Expansion,''
  Nucl.\ Phys.\ B {\bf 149} (1979) 285.
  doi:10.1016/0550-3213(79)90243-8
  
  \bibitem{Witten2}
  E.~Witten,
  ``Current Algebra Theorems for the U(1) Goldstone Boson,''
  Nucl.\ Phys.\ B {\bf 156} (1979) 269.
  doi:10.1016/0550-3213(79)90031-2
  
  \bibitem{Witten3}
  E.~Witten,
  ``Theta dependence in the large N limit of four-dimensional gauge theories,''
  Phys.\ Rev.\ Lett.\  {\bf 81} (1998) 2862
  doi:10.1103/PhysRevLett.81.2862
  [hep-th/9807109].
 
  
  \bibitem{CDG2}
  C.~G.~Callan, Jr., R.~F.~Dashen and D.~J.~Gross,
  ``Toward a Theory of the Strong Interactions,''
  Phys.\ Rev.\ D {\bf 17} (1978) 2717.
  doi:10.1103/PhysRevD.17.2717
  
 
  
  \bibitem{Veneziano}
  G.~Veneziano,
  ``U(1) Without Instantons,''
  Nucl.\ Phys.\ B {\bf 159} (1979) 213.
  doi:10.1016/0550-3213(79)90332-8
  
  \bibitem{VW}
  C.~Vafa and E.~Witten,
  ``Parity Conservation in QCD,''
  Phys.\ Rev.\ Lett.\  {\bf 53} (1984) 535.
  doi:10.1103/PhysRevLett.53.535
%  \bibitem{susskind}
%  Y. Aharonov and L. Susskind
%  "Charge superselection rule"
%Phys. Rev. 155, 1428 (1967)
\bibitem{WittenO}
  E.~Witten,
  ``Some Properties of O(32) Superstrings,''
  Phys.\ Lett.\  {\bf 149B} (1984) 351.
  doi:10.1016/0370-2693(84)90422-2
  \bibitem{WittenW}
  X.~G.~Wen and E.~Witten,
  ``World Sheet Instantons and the {Peccei-Quinn} Symmetry,''
  Phys.\ Lett.\  {\bf 166B} (1986) 397.
  doi:10.1016/0370-2693(86)91587-X
  \bibitem{Wittenaxion}
  P.~Svrcek and E.~Witten,
  ``Axions In String Theory,''
  JHEP {\bf 0606} (2006) 051
  doi:10.1088/1126-6708/2006/06/051
  [hep-th/0605206].
  \bibitem{Banks}
  T.~Banks, M.~Dine, P.~J.~Fox and E.~Gorbatov,
  ``On the possibility of large axion decay constants,''
  JCAP {\bf 0306} (2003) 001
  doi:10.1088/1475-7516/2003/06/001
  [hep-th/0303252].

\end{thebibliography}
\end{document}